\documentclass[journal]{IEEEtran}
\usepackage{amsmath}
\usepackage[short]{optidef}
\usepackage{xcolor}
\usepackage{comment}
\usepackage{algorithmic}  
\usepackage{algorithm}
\usepackage{amsmath}
\usepackage{amssymb}
\usepackage{graphicx}
\usepackage{epstopdf}
\usepackage{multirow}
\usepackage{booktabs}
\usepackage{cite}
\usepackage{color}
\usepackage{tabularx}
\usepackage{mathtools}
\usepackage{float}
\usepackage{enumerate}
\usepackage{algorithm,amsmath,amssymb,amsfonts}
\usepackage{subcaption}
\begin{document}
\font\myfont=cmr12 at 18pt
\title{\myfont Joint Computation Offloading and Target Tracking in Integrated Sensing and Communication Enabled UAV Networks}

\author{Trinh Van Chien, Mai Dinh Cong, Nguyen Cong Luong, Tri Nhu Do, Dong In Kim, and Symeon Chatzinotas \vspace{-1cm}
\thanks{This research was supported in part by the National Research Foundation of Korea (NRF) Grant funded by the Korean Government (MSIT) under Grant 2021R1A2C2007638, and in part by Vietnam
National Foundation for Science and Technology Development (NAFOSTED)
under grant number 102.02-2023.43. This research is funded by Hanoi University of Science and Technology (HUST) under project number T2022-TT-001 for Trinh Van Chien. \textit{(Corresponding author: Nguyen Cong Luong)}}
\thanks{Trinh Van Chien and Mai Dinh Cong are with the School of Information and Communication Technology, Hanoi University of Science and Technology, Vietnam (e-mails: chientv@soict.hust.edu.vn and  cong.md204521@sis.hust.edu.vn). Nguyen Cong Luong is with the Faculty of Computer Science
Phenikaa University,
Hanoi, Vietnam (e-mail: luong.nguyencong@phenikaa-uni.edu.vn). 
Tri Nhu Do is with the Department of Electrical Engineering, Polytechnique Montr\'{e}al, Montr\'{e}al, QC H3T 1J4, Canada (e-mail: tri-nhu.do@polymtl.ca). Dong In Kim is with the Department of Electrical and Computer Engineering, Sungkyunkwan University, Suwon 16419, South Korea (e-mail:dikim@skku.ac.kr). Symeon Chatzinotas is with the Interdisciplinary Centre for Security, Reliability and Trust, University of Luxembourg (e-mail: symeon.chatzinotas@uni.lu).}
}

\maketitle

\begin{abstract}
In this paper, we investigate a joint computation offloading and target tracking in Integrated Sensing and Communication (ISAC)-enabled unmanned
aerial vehicle (UAV) network. Therein, the UAV has a computing task that is partially offloaded to the ground UE for execution. Meanwhile, the UAV uses the offloading bit sequence to estimate the velocity of a ground target based on an autocorrelation function. The performance of the velocity estimation that is represented by Cramer-Rao lower bound (CRB) depends on the length of the offloading bit sequence and the UAV's location. Thus, we jointly optimize the task size for offloading and the UAV's location to minimize the overall computation latency and the CRB of the mean square error for velocity estimation subject to the UAV's budget. The problem is non-convex, and we propose a genetic algorithm to solve it. Simulation results are provided to demonstrate the effectiveness of the proposed algorithm.
\end{abstract}


\section{Introduction}
 Owing to unique advantages of high altitude operations, high mobility, and strong air-ground sight (LoS) links, unmanned aerial vehicles (UAVs) have been proposed as Internet of Thing (IoT) devices to effectively collect image data, e.g., for virtual reality and Metaverse systems. However, the UAVs have constraints in computing energy and networking resources. Therefore, it is challenging to handle intensive tasks of data processing such as image feature extraction. To address the issue, edge computing is still an effective solution that allows the UAV to offload a part of the computing task to ground edge servers or ground user equipment (UE)~\cite{messous2019game}. It is a fact that since the UAV knows the offloading bit sequence in terms of the length and their bit values, the UAV can use the bit sequence to detect or track ground targets. The UAV can do this since it can be equipped with integrated sensing and communication (ISAC) capability. ISAC allows a device to use the same hardware, energy, and spectrum resources to simultaneously perform the communication function to transmit its data and radar function to detect/track a target. In this work, we assume that the UAV uses the communication signal for the radar function. Particularly, the communication function allows the UAV to transmit the bit sequence of the task to the UE while the sensing function helps it to perform the estimation of the target's parameter by using the bit sequence. However, the radar tracking performance depends on the length of the offloading bit sequence. Therefore, the offloading bit sequence needs to be optimized. 

Based on the above observations, in this work, we investigate the joint computation offloading and target tracking in an ISAC-enabled UAV network. The proposed system consists of a UAV, a ground UE, and a ground target that the UAV tracks. The UAV has a computing task, and it partially offloads the task to the UE. While offloading this task to the UE, the UAV tracks the target on the ground to measure the velocity of the target by using an autocorrelation function on the offloading bits. To evaluate the precision of estimating the target's velocity, we leverage the Cramer-Rao lower bound (CRB) on the mean square error of velocity estimation as presented in \cite{kumari2017ieee}. In general, the higher size of the offloaded part of the task increases the target tracking performance, i.e., lower CRB, but this may increase the offloading latency as well as the offloading cost. Then, we optimize the part of the task to be performed locally at the UAV and its location to minimize the overall computation latency and the CRB subject to the UAV budget. The optimization problem is non-convex, and we propose a genetic algorithm to solve it. Simulation results are provided to demonstrate the effectiveness of the proposed algorithm. 

To the best knowledge of the authors, this is the first paper that aims to investigate joint computation offloading and target tracking in ISAC-enabled UAV networks. Indeed, there are existing works investigating the task offloading from the UAV to a ground base station. Such work can be found in~\cite{messous2019game} that aims to optimize the offloading decisions of the UAV to minimize energy consumption, delay time, and communication cost. However, target tracking is not considered. There are also several works investigating ISAC-enabled UAVs including the use of communication signals for radar, but edge computing is not considered. Particularly, in~\cite{qin2023deep}, multiple UAVs serve as mobile ISAC platforms to sense and communicate with on-ground target users. The problem is to maximize the radar sensing, communication throughput, and energy efficiency of the network. This can be achieved by optimizing the UAVs' trajectory planning, user association, and power allocation. In~\cite{wang2020constrained}, the author proposed an ISAC-equipped UAV network in which multiple UAVs simultaneously provide downlink communications to ground mobile users while cooperatively performing radar functions to detect the location of ground targets. By jointly optimizing the UAV's locations, user association, and transmit power, the work aims to maximize a network utility function, including the sum rate over the UAVs while guaranteeing the CRB of the location estimation. Different from~\cite{qin2023deep} and \cite{wang2020constrained}, the authors in~\cite{deng2023beamforming} investigated a UAV-enabled ISAC system that allows the UAV to schedule the sensing function and communication function according to practical application requirements. This helps to significantly reduce network resources. A recent work~\cite{9729765} investigated joint communication, radar sensing, and edge computing with multiple ISAC-equipped UE. However, our work is different from this work where the UE collects the sensing data via the radar function and then offloads a portion of the data via its radar function to a base station for processing.  

\vspace{-0.25cm}
\section{System Model}
\label{sec:system}

\begin{figure}[t]
	\centering
	\begin{minipage}[t]{0.24\textwidth}
	\includegraphics[trim=0.0cm 0.0cm 0.00cm 0.1cm, clip=true, width=1.6in]{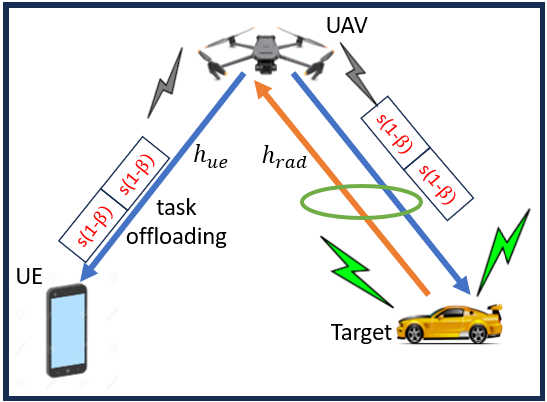}
		\subcaption{}
	\end{minipage}
	\begin{minipage}[t]{0.24\textwidth}
\includegraphics[ width=1.7in]{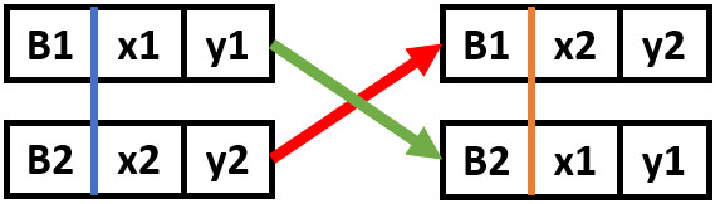}
		\subcaption{}
	\end{minipage}
	\caption{(a) Computation offloading and target tracking in an
ISAC-enabled UAV Network (b) Illustration of one-point crossover.}
	\label{UAV_MEC}
 \vspace{-0.25cm}
\end{figure}
We consider an
ISAC-enabled UAV network as shown in Fig.~\ref{UAV_MEC}(a). The network consists of a ground user equipment (UE), a ground target, and an ISAC-equipped unmanned aerial vehicle (UAV). The UAV has a computing task, and it can partially offload the task to UE. We define the computation task of the UAV as a tuple of $(w,s)$, where $s$ is the task size in bit and $w$ represents the number of CPU cycles required to compute one bit. We denote $\beta, \beta \in [0,1)$ as the task fraction that the UAV locally processes, and thus $1-\beta$ as the part of the task offloaded to the UE. 

While offloading this task to the UE, the UAV uses the offloading bits to perform tracking the target on the ground to measure parameter of the target. In this work, we assume that the UAV is willing to measure the velocity of the target. Note that different parameters such as target distance can be considered. The UAV can use the offloading bits for measuring the parameter of the target due to the following reasons. First, the UAV can add one copy of the offloading bit sequence, i.e., $(1-\beta)s$ bits, to the offloading bits. Second, the UAV uses an omnidirectional antenna such that both the UE and the target receive the offloaded signals. Then, the signal reaching the target is reflected to the UAV. Third, the UAV performs an autocorrelation operation between the received signal and its delayed version. Finally, the velocity of the target is estimated when the value of the autocorrelation function is maximum. 
\subsection{Channel Model}
We denote $[x,y, H]$ as the coordinate of the UAV, where $H$ is fixed. The horizontal coordinates of the ground user and the target are is fixed at $[u,v]$, and that of the target is $[x_{\rm{tar}}, y_{\rm{tar}}]$. Because of the high altitude of the UAV, we assume that the communication link and radar sensing between UE-UAV, UAV and target are dominated by the LoS links where the channel quality depends mainly on the distance between them. In radar sensing, the localization is based on the round-trip delay estimation of echo signals directly reflected from the target, and there are no obstructions during the reflection. The free-space path loss model is used for both the communication channel and the radar channel. The communication channel from the UE to the UAV, denoted by $h_{ue}$, is given by\footnote{The fluctuation loss is aligned with the Swerling I model.} 
\begin{equation}
    h_{ue} = \frac{h_0}{d_{ue}^2},
\end{equation}
where $h_0$ denotes the channel power at the reference distance
$d_0 = 1$ m, and $d_{ue}$ is the Euclidian distance between the UE and UAV.
We denote $h_{\rm{rad}}$ as the radar channel, which is 
\begin{equation}
    h_{\rm{rad}}= \frac{h_0}{d_{\rm{tar}}^2 L_{\rm{PL}}},
\end{equation}
where $d_{\rm{tar}}$ is the distance between the UAV and the target, $L_{\rm{PL}}$ is the path-loss factor defined by\cite{11222222}
\begin{equation}
    L_{\rm{PL}} = \frac{4\pi d^2_{\rm{tar}}}{\sigma_{\rm{RCS}}},
\end{equation}
where $\sigma_{\mathrm{RCS}}$ is the radar cross section of the target.

\vspace{-0.25cm}
\subsection{Offloading Time}
We denote $\xi^{\rm{uav}}$ as the available computation resource of the UAV. The delay caused by the local computation at the UAV is denoted by $t_{\rm{comp}}^{\rm{uav}}$, which is determined by~\cite{hoa2023deep}
\begin{equation}
    t_{\rm{comp}}^{\rm{uav}} = \frac{ws\beta}{\xi^{\rm{uav}}}.
\end{equation}
Since the UAV adds a copy of $s(1-\beta)$ bits to the offloading bit sequence of size of $s(1-\beta)$, the delay caused by the task communication, denoted by $t_{\rm{com}}$, is define by
\begin{equation}
    t_{\rm{com}} = \frac{2s{(1-\beta)}}{r},
\end{equation}	
where $r$ is the average transmission rate from the UAV to the UE, which is calculated by
\begin{equation}
    r = \mathit{B}\log_2 \left(1+ \frac{P}{BN_0}h_{\rm{ue}} \right),
\end{equation}
where $B$ and $P$ are the system bandwidth and transmit power of UAV, respectively, $N_0$ is the power spectral density (PSD). When the UE receives $2(1-\beta)s$ bits, it extracts the first $(1-\beta)s$ bits for the execution. 
We denote $\xi^{\rm{ue}}$ as the available computation resource of the UE. The delay caused by the computation time at the UE, denoted by $t_{\rm{comp}}^{\rm{ue}}$, is
\begin{equation}
    t_{\rm{comp}}^{\rm{ue}} = \frac{ws(1-\beta)}{\xi^{\rm{ue}}}.
\end{equation}
The offloading time is defined by
\begin{equation}
    t_{\rm{off}} = t_{\rm{com}}+t_{\rm{comp}}^{\rm{ue}}.
\end{equation}
Since the local computation at the UAV and the task offloading are executed simultaneously, the overall latency is
\begin{equation}
    t_{\rm{total}} = \max(t_{\rm{comp}}^{\rm{uav}}, t_{\rm{off}}).
\end{equation}

\subsection{Offloading Cost}
\subsubsection{Communication cost}
During the task offloading, the UAV consumes the bandwidth and energy to transmit $2(1-\beta)s$ bits to the UE. Thus, the communication cost, denoted by $c_{\rm{com}}$, is 
\begin{equation}
    c_{\rm{com}} = a2({1-\beta})s + \chi_1 Pt_{\rm{com}},
\end{equation}
where $a$ is the bandwidth price for transmitting one bit, $\chi_1$ denotes the price for consuming one unit of energy when transmitting.
\subsubsection{Computation Cost}
The UAV also needs to pay the UE that consumes its computing resource and energy resource for executing $(1-\beta)s$ bits is 
\begin{equation}
    c_{\rm{comp}} = ws({1-\beta})b +  \chi_2 ws(1-\beta)c, 
\end{equation}
where $b$ is the price for consuming one CPU cycle, $\chi_2$ denotes the energy price for consuming one unit of energy at the UE, and $c$ represents the CPU energy consumption required to implement one CPU cycle.

Finally, the total offloading cost, denoted by $c_{\rm{total}}$, is
\begin{equation}
    c_{\rm{total}} = c_{\rm{com}}+c_{\rm{comp}}.
\end{equation}


\subsection{Target Tracking}
The UAV estimates the velocity based on the autocorrelation technique. This technique implements the correlation of the offloading signal with a delayed copy of itself as a function of delay. Therefore, the UAV needs to send a sequence consisting of two identical bit sub-sequences, each of which has a size of $s(1-\beta)$. As such, the bit sequence has the size of $2s(1-\beta)$ bits. We denote $y[l]$ as the baseband signal that the UAV receives due to the reflection from the target. Here, $l$ refers to the sampling time index, that is normalized by a sampling duration of $T_s$ with $T_s=1/B$. Then, the autocorrelation function is implemented over a window of $L$ samples by~\cite{kumari2017ieee}
\begin{equation}
    R[l] = \frac{{\sum_{n=0}^{L-1} y^*[l-n]y[l-n-L]}}{\sqrt{{\sum_{n=0}^{L-1} |y^*[l-n]|^2}} \sqrt{ \sum_{n=0}^{L-1} |y^[l-n-L]|^2}},
    \label{auto-corr}
\end{equation}
where $L$ is the window size of the autocorrelation function and $y^*[l-n]$ is the conjugate complex function of $y[l-n]$. In general, the high value of $L$ leads to a higher estimation of the target velocity. However, $L$ must be smaller or equal to $s(1-\beta)$, i.e., $ L \leq s(1-\beta)$. The reason is that the autocorrelation function is applied to the two identical sub-sequences, each of which has the length of $s(1-\beta)$. Therefore,  we set $L = s(1-\beta)$. The beginning of the first sub-sequence, denoted by $\hat{l}$, is estimated by~
\begin{equation}
    \hat{l} = \arg \max_{l}  R[l].
\end{equation}
Then, the velocity of the target is determined by taking the angle, represented by $\angle$, of $R[l=\hat{l}]$ as follows~\cite{kumari2017ieee}
\begin{equation}
    \nu_0 =\frac{\angle \left(R[\hat{l}] \right)}{2\pi T_D}, 
\end{equation}
where $T_D= LT_s$. In general, the accuracy of estimation of $\nu_0$ depends on the signal-to-noise ratio (SNR) of the signal received at the UAV, the bandwidth, and the length of the sub-sequence. To show this relationship and to be convenient for network optimization, we use the CRB as a lower bound on the variance of an unbiased estimator. The CRB is considered to be the lower bound on the mean square error of the velocity estimation. For the autocorrelation function given in \eqref{auto-corr}, the CRB is given by~\cite{kumari2017ieee}

\begin{equation}
    \texttt{CRB}_v = \frac{6\lambda^2}{16(\pi^2)({1-\beta})^3 s^3T_s^2 \gamma_r},
\end{equation}
where
$T_s \approx \frac{1}{B}$ is the symbol duration related to the system bandwidth,
$\lambda$ is the signal wavelength, and
$\gamma_{\rm{rad}}$ is the SNR of the signal received at the radar defined by
\begin{equation}
    \gamma_{\rm{rad}} = \frac{Ph_{\rm{rad}}}{BN_0}.
    \label{SNR_rad}
\end{equation}


\section{Problem Formulation and Genetic Algorithm}
In this section, we first formulate the optimization problem, and then we propose the genetic algorithms to solve it.
\subsection{Problem Formulation}
We aim to minimize the overall computation latency and minimize the CRB of the MSE of the target velocity estimation subject to the UAV's budget. For this, we optimize the task-splitting factor of $\beta$ and the locations of the UAV. Therefore, the optimization problem is mathematically formulated as
\begin{equation}
\label{opt-pro}
\begin{aligned}
& \min_{\beta,x,y} && z= w_1t_{\rm{total}} +w_2\texttt{CRB}_v \\
\textrm{s.t.\; }& \textrm{C1 :}  && 0 \leq \beta \leq \beta_{\rm{max}}, \\
& \textrm{C2 :}&& c_{\rm{total}} \leq c_{\rm{budget}},\\
\end{aligned}
\end{equation}
where $w_1$ and $w_2$ are the weights associated with the overall computation latency and the CRB, respectively. The values of $w_1$ and $w_2$ are selected depending on the priority levels of the sub-objectives of the UAV. Then, they can be determined via experiment observation. $\beta_{\rm{max}}$ represents the upper limit of $\beta$ and must be less than one. The constraint in (C2) ensures that the total offloading cost does not exceed the budget $c_{\rm{budget}}$ of the UAV, which is the maximum amount of money the UAV can pay. The optimization problem in \eqref{opt-pro} is non-convex and is challenging to be solved. Particularly, optimization algorithms such as exhaustive search cannot be used since the location and the task size $\beta$ are continuous variables. We thus propose a genetic algorithm to solve it as presented in the next section.\footnote{Due to the inherent non-convexity, the GA-based method cannot guarantee the global or local optimum for problem~\eqref{opt-pro}. Nonetheless,  we can at least know that the global optimum exists and the proposed meta-heuristic algorithm can obtain a sub-optimal solution in polynomial time.}

\subsection{Genetic Algorithm}
The algorithm starts by generating an initial population in which every chromosome satisfies the problem's constraints. Next, the algorithm computes the fitness of each chromosome, indicating its performance is the objective function value of the total latency $t_{\rm{total}}$ and the $\mathtt{CRB}$. Subsequently, two chromosomes are selected to serve as parents. Chromosomes with higher fitness have a higher probability of being selected as parents. Afterward, the crossover and mutation processes are performed using the two selected parents to generate two novel solutions. This process is repeated until all of the solutions in the new generation are created. After that, the tournament selection method is used to select new individuals for survival for the next generation. At this point, the old population is replaced with the new one. The algorithm iterates through multiple generations and terminates at a predefined number of iterations from previous ones. The output of the algorithm is the best solution to the problem
\subsubsection{Solution Representation}
An individual designated by a chromosome in a genetic algorithm represents a solution to the problem. The encoding of the solution consists of $\beta$, $x$, and $y$, which represent real-valued genes.
\subsubsection{Crossover Operation}
Fig.~\ref{UAV_MEC}(b) shows the crossover operation process. After two parents are selected, one point crossover is used to generate offspring with a probability of $P_c$. The algorithm randomly selects a point as a crossover point in a chromosome of the parent. Based on the crossover point, the genes on the left and right sides of the crossover point in the chromosome are exchanged, and two new chromosomes are generated. 

\subsubsection{Mutation Operation}
After the crossover operation is implemented, two chromosomes have been generated, and the algorithm applies the polynomial mutation operation with a defined index parameter of $\eta$. However, numerous studies suggest that the value of $\eta$ should fall within the range of $[20, 100]$ \cite{mutationOperator}. For each chromosome, a random number is selected in a range of $(0, 1)$. As the random number is less than a predefined probability $P_m$, the mutation is applied to this chromosome. For the given element $p$ in the chromosome with $p \in [a',b']$, where $a'$ and $b'$ are lower and upper bounds of the variable, a mutated value of $p'$ is created for a random number $u$ generated within $[0, 1]$ as follows:
\begin{equation}
p' = \begin{cases}
p + \bar{\delta}L(p - a') & \text{for } u \leq 0.5, \\
p + \bar{\delta}R(b' - p) & \text{for } u > 0.5,
\end{cases}
\end{equation}
where $\bar{\delta}L$ and $\bar{\delta}R$ are computed as follows
\begin{align}
&\bar{\delta}L = (2u)^{\frac{1}{1+\eta}} - 1, \text{for } u \leq 0.5,\\
&\bar{\delta}R = 1 - \left(2(1 - u)\right)^{\frac{1}{1+\eta}}, \text{for } u > 0.5.
\end{align}
\subsubsection{Survival Selection}
After generating a new set of individuals, a tournament selection method is applied to select a new population for the next generation. Two individuals are randomly selected each time, and the better individual is selected and added to the population. This process is repeated until the new population reaches the initial population size. The method helps to maintain balance and diversity in the population with low computational complexity.
\subsubsection{Computation Complexity}

\begin{algorithm}
\small
\caption{Proposed GA for ISAC-enabled UAV network}
\textbf{Begin}\\
\textbf{Input: $P_m$, $P_c$, $T$, $K$}
\begin{algorithmic}
\STATE \textbf{Initialization:} a population denoted by $\mathtt{K}$ with size of $K$.
\STATE Calculate the fitness value of each individual, and select the individual with the highest fitness as $S_{\rm{best}}$; its fitness value is denoted as $F_{\rm{best}}$. Please note that the fitness value is calculated as $ \frac{1}{z} $, where $ z $ represents the objective function defined in (18).

\FOR{$t = 1$ \TO $T$}
\STATE Initialize two empty populations denoted by $\mathtt{K}^{\rm{empty}}$ and $\mathtt{K}^{\rm{new}}$.
\FOR{$k = 1$ \TO $K$}
    \STATE Randomly select two individuals for crossover $R_1$ and $R_2$, the individual with higher fitness has a higher probability of being selected.
    \IF{$rand([0,1])<P_c$}
        \STATE Update $R_1$ and $R_2$ by Perform Crossover like Fig.~\ref{UAV_MEC}(b).
    \ENDIF
    \IF{$rand([0,1])<P_m$}
        \STATE Mutate the offspring according to (19);
    \ENDIF
    
    \STATE $\mathtt{K}^{\rm{empty}} \longleftarrow \mathtt{K}^{\rm{empty}} \cup {R_1}$ and $\mathtt{K}^{\rm{empty}} \longleftarrow \mathtt{K}^{\rm{empty}} \cup {R_2}$
\ENDFOR
\FOR{$k=1$ \TO $K$}
   \STATE $R_1$ = rand($\mathtt{K}^{\rm{empty}}$) and $R_2$ = rand($\mathtt{K}^{\rm{empty}}$)
   \IF{fitness($R_1$) $\geq$ fitness($R_2$)}
        \STATE $\mathtt{K}^{\rm{new}} \longleftarrow \mathtt{K}^{\rm{new}} \cup R_1$
    \ELSE
        \STATE  $\mathtt{K}^{\rm{new}} \longleftarrow \mathtt{K}^{\rm{new}} \cup R_2$
    \ENDIF
\ENDFOR
    \STATE $K \longleftarrow \mathtt{K}^{\rm{new}}$
    \STATE Compare the best individual $S'_{\rm{best}}$ and its fitness value $F'_{\rm{best}}$ from iteration $t$ with $S_{\rm{best}}$.
    \IF{$F'_{\rm{best}} > F_{\rm{best}}$}
        \STATE $S_{\rm{best}} \longleftarrow S'_{\rm{best}}$\\$F_{\rm{best}} \longleftarrow F'_{\rm{best}}$.
    \ENDIF
\ENDFOR
\STATE \textbf{Output:} $S_{\rm{best}}, F_{\rm{best}}$
\STATE \textbf{End}
\end{algorithmic}
\end{algorithm}
The time complexity of the proposed genetic algorithm depends on the
time complexity of the survival selection, crossover operation, and mutation operation. In the selection operation, to select a new generation, the time complexity is $\mathcal{O}(T K)$, where $T$ represents the number of iterations, and $K$ represents the number of individuals in the population. Meanwhile, the time complexity for crossover and mutation is simply $\mathcal{O}(TK)$. Thus, the overall time complexity of the algorithm is $\mathcal{O}(TK)$.

\section{Performance Evaluation}
\vspace{-0.1cm}
In this section, we present numerical findings to evaluate the performance of the proposed genetic algorithm. The system parameters are set as follows. The transmit power is $P_{\max} = 27$ dBm, the computing capacity of the UAV and UE are $\xi^{\rm{uav}} = 6\times 10^6$ cycles/s and $\xi^{\rm{ue}} = 5\times 10^6$ cycles/s, respectively. The UAV operates at an altitude of $60$ m. The system bandwidth is $B = 10^7$ Hz. For the network model, we assume that the UAV, UE, and the target all operate within a square area of $1000$ m $\times$ $1000$ m. Other parameters are presented in Table~\ref{table:parameters_CRN}. In the result figues, we name the proposed algorithm as proposed GA. To evaluate the effectiveness of the proposed GA, we introduce the proposed GA with fixed $\beta$ that uses the genetic algorithm to optimize only the location of the UAV and the proposed GA with fixed $(x,y)$ that uses the genetic algorithm to only optimize $\beta$.

\begin{figure}[t]
	\centering
	\begin{minipage}[t]{0.24\textwidth}
	\includegraphics[trim=1.2cm 0.1cm 0.2cm 1.6cm, clip=true, width=1.8in,height = 1.05in]{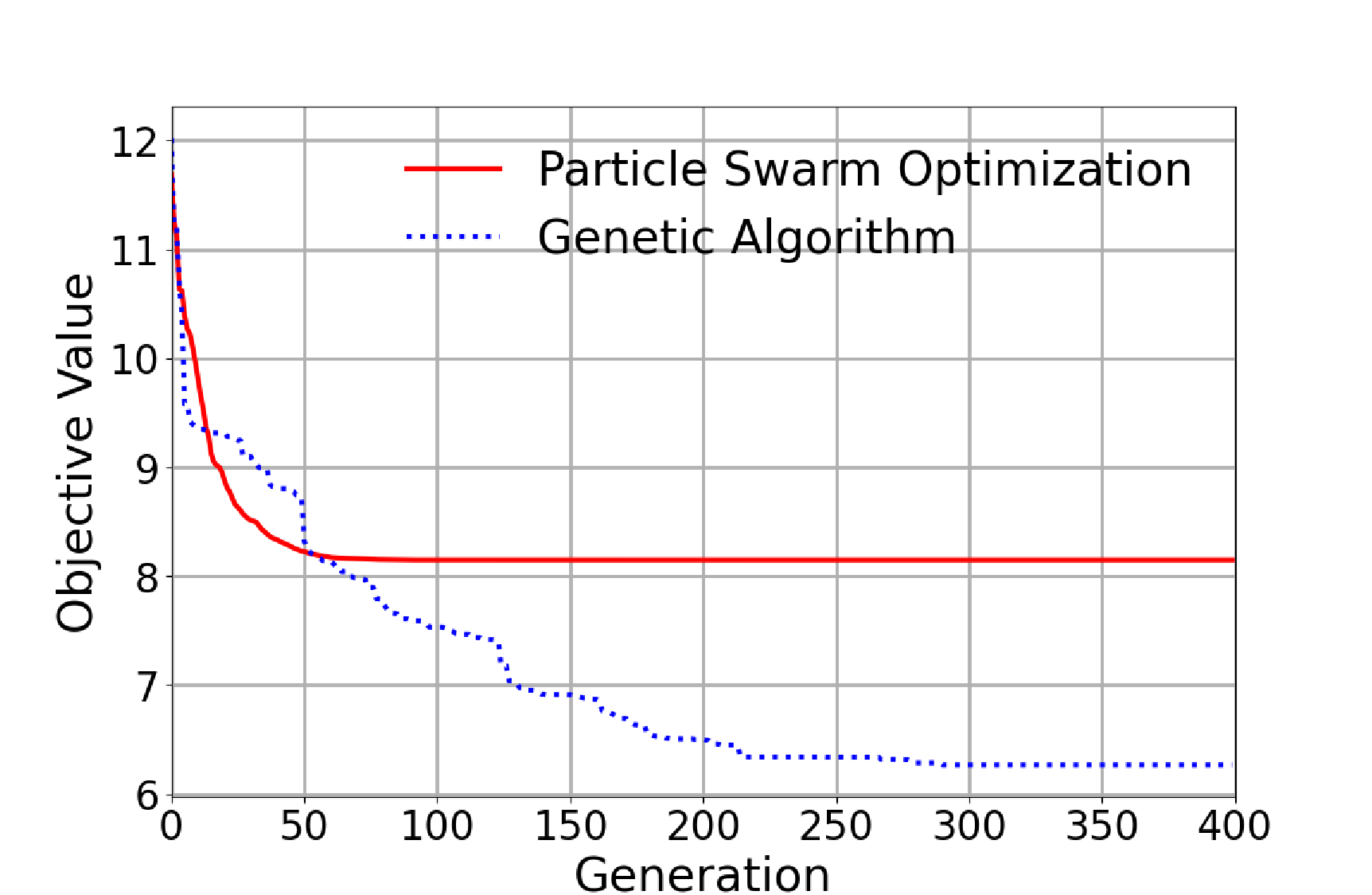}
		\subcaption{}
	\end{minipage}
	\begin{minipage}[t]{0.24\textwidth}
\includegraphics[ width=1.9in]{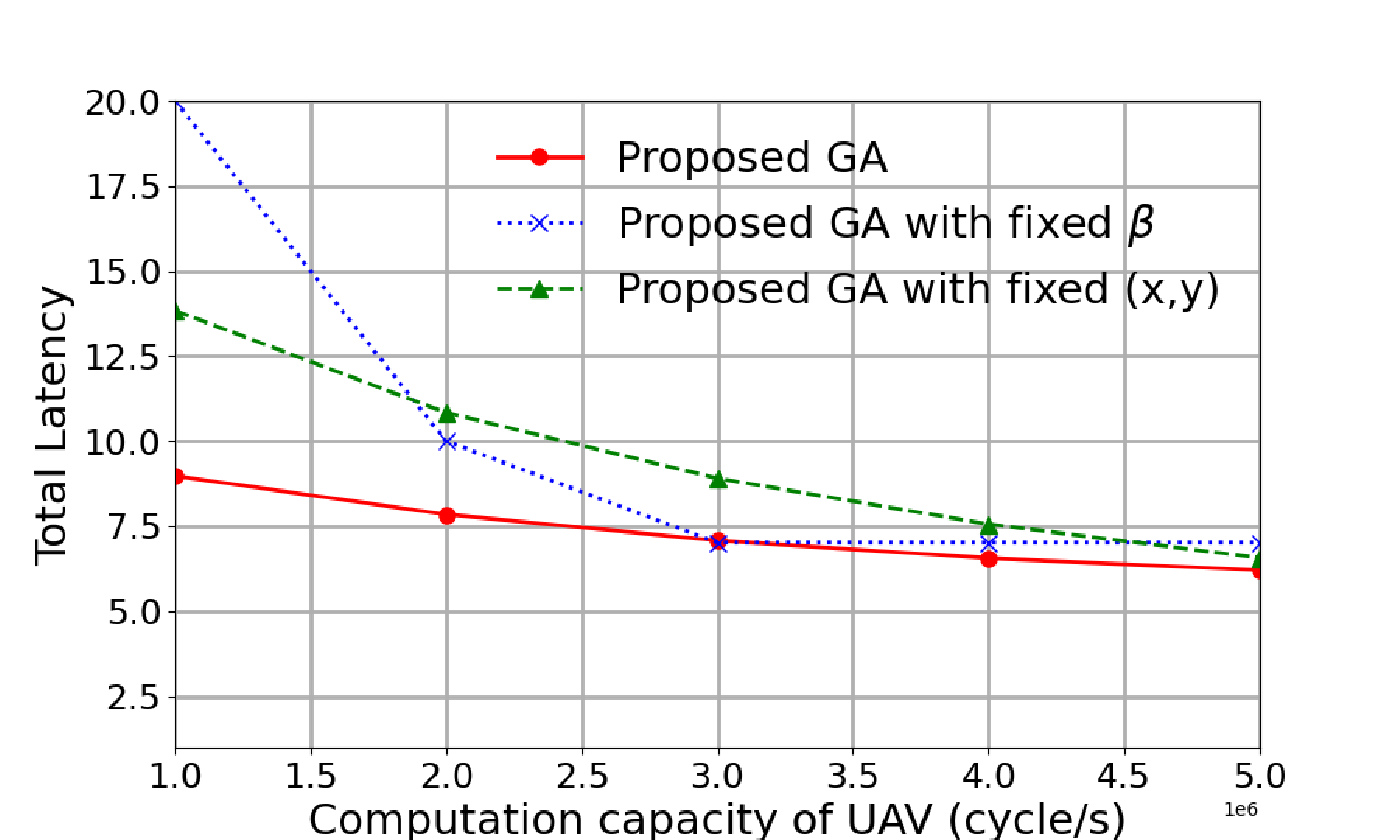}
		\subcaption{}
	\end{minipage}
	\caption{(a) Convergence of algorithms (b) Total latency versus the UAV's computation capacity.}
	\label{Convergence}
 \vspace{-0.25cm}
\end{figure}

\begin{table}[h]
\caption{\small Simulation parameters}
\label{table:parameters_CRN}
\footnotesize
\centering
\begin{tabular}{lc|lc}
\hline\hline
{\em Parameter} & {\em Value} & {\em Parameter} & {\em Value}\\ [0.5ex]
\hline
$s$ & $5\times10^{6}$ bits  & $w$ & $10$ cycle/bit \\
\hline
$x_t$ & $460$ & $y_t$ & $290$ \\
\hline
$u$ & $100$& $v$ & $120$\\
\hline
$a$ & $50$& $b$ & $10$\\
\hline
$n$ & $2$& $\sigma_{\rm{RCS}}$ & $0.1$ $\mathrm{m}^2$\\
\hline
$\lambda$ & $0.03$m & $N_0$ & $10^{-17}$ Watt/Hz\\
\hline
$w_1$ & $1$& $w_2$ & $40$\\
\hline
$P_m$ & $0.15$ & $P_c$ & $0.8$\\
\hline
$T$ & $400$ & $K$ & $20$\\
\hline
\end{tabular}
\label{table:parameters}
\end{table}
To evaluate the effectiveness of the proposed GA, we introduce the Particle Swarm Optimization (PSO) as a baseline algorithm. As shown in Fig.~\ref{Convergence}(a), the convergence speed of the PSO is faster than that of the proposed GA since the PSO attempts to prioritize exploitation over exploration. Particularly, the particles adjust their positions based on their own best-known position within the swarm as well as their previous best-known position, thereby leading to a more focused search around promising solutions. On the other hand, GA involves crossover and mutation operations, promoting more exploration of the search space. However, the performance obtained by the proposed GA is higher than that obtained by the PSO. The reason is that in the investigated scenario, the proposed GA can explore a larger solution space, making it easier to achieve better results.


It's worth discussing how the computing resource $\xi^{\rm{uav}}$ of the UAV impacts the total latency achieved by the algorithms. As illustrated in Fig.~\ref{Convergence}(b), the increase of $\xi^{\rm{uav}}$ results in a reduction of the total latency achieved by the algorithms. This is obvious since the UAV is able to finish the offloaded task earlier.
However, with the GA algorithm with fixed $\beta$, the total latency first decreases and then keeps stable as $\xi^{\rm{uav}}$ increases, i.e., $\xi^{\rm{uav}} \geq 3\times10^6$. This is because the total latency is primarily determined by the maximum of the offloading time to the UE and the computation time at the UAV. When the computation capacity of the UAV increases, the computation time at the UAV decreases while the offloading time keeps constant. 
Thus, the total latency remains unchanged. As such, selecting the most appropriate percentage of task offloading is very important to minimize the total latency.

\begin{figure}[t]
	\centering
	\begin{minipage}[t]{0.24\textwidth}
	\includegraphics[trim=1.5cm 0.1cm 0.2cm 1.6cm, clip=true, width=1.8in]{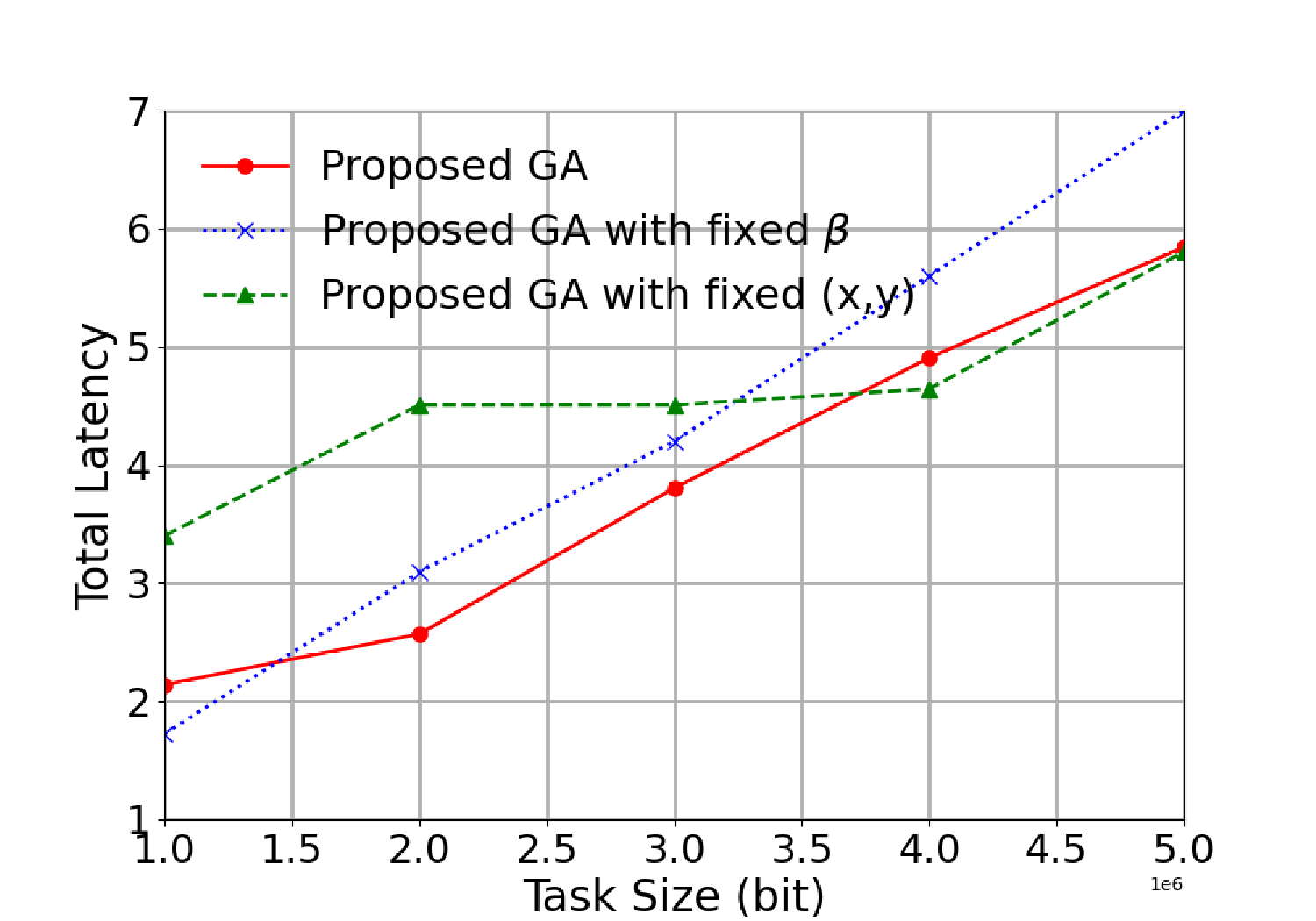}
		\subcaption{}
	\end{minipage}
	\begin{minipage}[t]{0.24\textwidth}
\includegraphics[ width=1.9in]{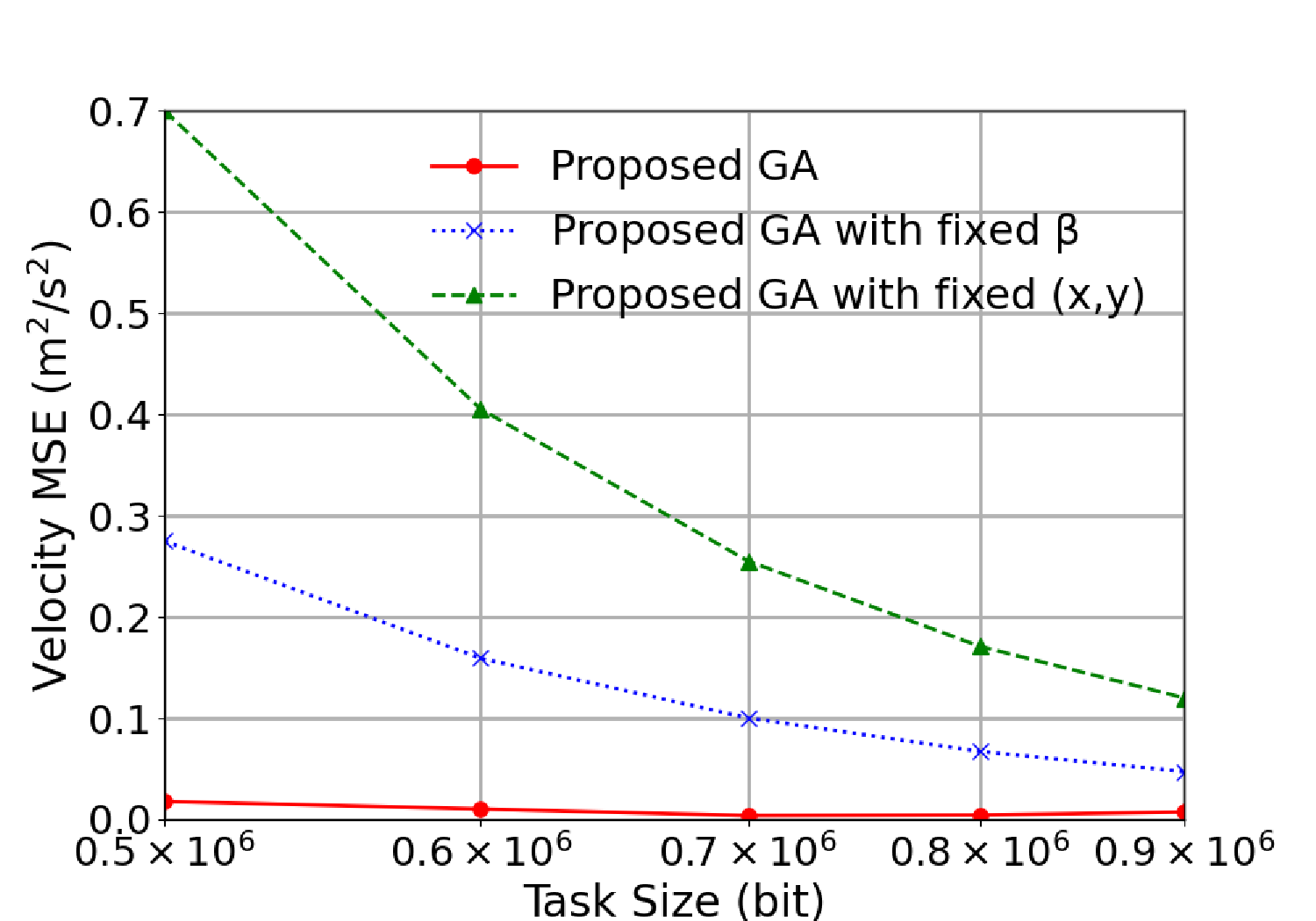}
		\subcaption{}
	\end{minipage}
	\caption{(a) Total latency and (b) MSE of the velocity estimation versus task size.}
	\label{task_size}
\end{figure}

Finally, the UAV may have tasks with different sizes, and thus we discuss how the task size impacts the overall computing latency and the radar performance, i.e., the CRB. As illustrated in Figs.~\ref{task_size}(a), the total latency increases as the task size increases. The reason is that the tasks with larger sizes increase the transmission latency and computing latency at both the UAV and UE. Meanwhile, as shown in Figs.~\ref{task_size}(b), as the task size increases, the CRB decreases meaning that the velocity estimation error decreases. 

\section{Conclusions}\label{sec:conclusion}
We have addressed the joint computation offloading and target tracking in an ISAC-enabled UAV network. The network allows the UAV to offload a part of its task to the ground UE while estimating the velocity of the ground target. The computation latency and the target tracking performance both depend on the task size to be offloaded and the UAV's location. Thus, we have formulated the optimization problem to optimize the task size for offloading and the UAV's location to minimize the computation latency and the CRB of the MSE of the velocity estimation error. We have developed the genetic algorithm to solve the optimization problem. Simulation results clearly show that the proposed algorithm outperforms the baseline schemes in terms of computation latency and CRB. In the future of work, we consider a general scenario including multiple UAVs that offload their tasks to multiple UEs while tracking multiple targets. Adapting the offloading bit length according to the relative distances between the UAVs and targets will be also an interesting work. In addition, clutter interference issue will need to be investigated.

\bibliographystyle{IEEEtran}
\bibliography{ref}
\end{document}